\documentclass[aps,prl,amssymb,amsmath,twocolumn,floatfix]{revtex4}
\usepackage{graphicx}
\usepackage{amsmath}
\graphicspath{{figures}{figures/snapshots}{figures/clusters}}

\newcommand{\vx}{\mathbf{x}}
\newcommand{\vy}{\mathbf{y}}

\begin{document}
\title{Mean-field interactions in evolutionary spatial games}
\author{Dmitriy Antonov$^{1}$}
\author{Evgeni Burovski$^{1}$}
\author{Lev Shchur$^{1, 2}$}
\affiliation{$^1$ HSE University, 101000 Moscow, Russia}
\affiliation{$^2$ Landau Institute for Theoretical Physics, 142432 Chernogolovka, Russia}

\begin{abstract}
We introduce a mean-field term to an evolutionary spatial game model. Namely, we consider the game of Nowak and May, based on the Prisoner's dilemma, and augment the game rules by a self-consistent mean-field term. This way, an agent operates based on local information from its neighbors and non-local information via the mean-field coupling. We simulate the model and construct the steady-state phase diagram, which shows significant new features due to the mean-field term: while for the game of Nowak and May, steady states are characterized by a constant mean density of cooperators, the mean-field game contains steady states with a continuous dependence of the density on the payoff parameter. Moreover, the mean-field term changes the nature of transitions from discontinuous jumps in the steady-state density to jumps in the first derivative. 
The main effects are observed for stationary steady states, which are parametrically close to chaotic states: the mean-field coupling drives such stationary states into spatial chaos.
Our approach can be readily generalized to a broad class of spatial evolutionary games with deterministic and stochastic decision rules.
\end{abstract}

\maketitle

\textit{Introduction---} 
The mean-field approximation, initially devised in 1907 in the context of (ferro)magnetic properties of materials~\cite{Weiss1907}, is since playing a truly central role in a variety of branches of physics, ranging from superconductivity~\cite{Sordi2012} to ferromagnetism of disordered metals~\cite{Andreev2004}; from ultracold quantum gases~\cite{Aikawa2012} to spin glasses~\cite{Gross1985}, to name just a few.
The main idea---that a behavior of a collection of a macroscopic number of strongly interacting components can be reduced to a single-particle problem in a self-consistent external potential---proves fruitful for multiple research fields outside of traditional physics: for instance, describing spatially extended networks~\cite{Bonamassa2019}, droplet traffic~\cite{Engl2005}, traffic flow~\cite{Schreckenberg1995}, and social dilemmas~\cite{Lin2019}.

In this Letter, we demonstrate how mean-field ideas can be applied to a  system that does not have an explicit representation in the language of statistical mechanics.
Instead, it is a deterministic dynamical system with a steady-state, i.e., a long-lived state with a well-defined ``thermodynamic limit'' of the infinite system size. We start with the completely deterministic evolutionary game model of Nowak and May~\cite{Nowak1992} and add a novel ingredient, the coupling to the instantaneous mean density of cooperators.  The modification does not change the deterministic nature of the process in contrast with the previous research~(see, f.e.,~\cite{Perc2017, Perc2021, Javarone2016, Javarone2016_memory, Javarone2018}) in which the stochastic nature of modifications obviously opens the possibility to using methods of statistical physics. Once again, our model is deterministic (specifically, there is no temperature, no fluctuations, and no random forces).
The introduction of the mean-field interaction leads to a nontrivial diagram of steady states and opens new directions for possible applications of structured multi-agent systems. Our approach can be generalized to any deterministic or stochastic spatial evolutionary game.

Studying the dynamics of ensembles of agents is a common theme in biology, economics,
and social sciences. Collective behavior emerges through repeated pairwise
interactions between individuals, where each agent acts to maximize its fitness.
The evolutionary game theory approach model elementary interactions as
game-theoretic contests between agents having varying strategies~\cite{Smith1982}. 
For well-mixed populations---in physics language, this corresponds to a
mean-field approximation---the macroscopic description proceeds via the master
equation for the time evolution of populations of strategies, the so-called
replicator equation~\cite{Weibull1995, Hofbauer2008}.

One limitation of the replicator equation approach is that it does not allow for \emph{structured} populations, where the interaction range is restricted to a local neighborhood. Such populations are known to exhibit long-time behavior beyond standard replicator dynamics~\cite{Hauert2005}. Various local arrangements have been considered: following the pioneering work~\cite{Nowak1992}, noisy dynamics were studied on regular, decorated, and frustrated two-dimensional lattices~\cite{Szolnoki2005, Hauert2005, Szolnoki2017} and coevolving random networks~\cite{Szolnoki2009}. Deterministic imitation dynamics have been investigated on regular square~\cite{Vainstein2014, Kolotev2018} and triangular lattices~\cite{Burovski2019}, on a simple cubic lattice 3D~\cite{Moskalenko2020}, and diluted 2D lattices~\cite{Vainstein2001, Vainstein2014}.

In the Letter, we construct an evolutionary game model which features both kinds of effects: purely local interactions with nearest neighbors, \emph{and} a self-consistent, mean-field-type coupling to the order parameter. We start with the classic game of Nowak and May (NM)~\cite{Nowak1992,Nowak1993}, where the order parameter can be chosen as the mean density of cooperators $\left< f_c\right>$ in the steady-state regime of evolution. The mean-field term is the coupling to the instantaneous density of cooperators, $f_c(t)$, which fluctuates due to the game dynamics and has a well-defined mean value in the steady-state regime. The addition of the MF term to the decision-making rule is quite realistic. It can be interpreted as the influence of the ``averaged strategy of society'' or a mass-media influence that reflects and drives society's opinions. In the limit of vanishing mean-field (MF) term, the phase diagram
reproduces known results.

We note that our approach is very different from what is known as mean-field games in mathematical economics~\cite{LasryLions2007}: the latter is a class of stochastic differential games (which, in turn, can be mapped onto the nonlinear Schroedinger equation~\cite{Swiecicki2016}). In our work, we only consider discrete deterministic games based on the Prisoner's dilemma.

\textit{The Prisoner's dilemma.---} Consider two agents, $\alpha$ and $\beta$, which interact via the rules of the Prisoner's dilemma game. An agent is characterized by a \emph{strategy}, which takes one of two values: cooperate, $\mathcal{C}$, or defect, $\mathcal{D}$.  In an ``interaction'', agents receive payoffs which depend on their strategies~\cite{Axelrod-1981}. Encoding strategies of an agent by length-two vectors, $\vec{s}$, such that $\mathcal{C} {=} (1, 0)^T$ and $\mathcal{D} {=} (0, 1)^T$, the payoff of an agent $\alpha$ is $P_\alpha {=}  \vec{s}_\alpha^{\,T} \widehat{H} \vec{s}_\beta$, and the payoff of the agent $\beta$ is $P_\beta{=}  \vec{s}_\beta^{\,T} \widehat{H} \vec{s}_\alpha$, where $\widehat{H}$ is the payoff matrix, and $\vec{s}_\alpha$ and $\vec{s}_\beta$ are the states of agents $\alpha$ and $\beta$, respectively. Following Ref.~\cite{Nowak1992}, we take
\begin{equation}
\widehat{H} = %
\begin{pmatrix}
1 & 0 \\
b & 0
\end{pmatrix} \;.
\label{eqn:payoff_matrix_NM}
\end{equation}
Essentially, Eq.~\eqref{eqn:payoff_matrix_NM} means that in a $\mathcal{C}$-$\mathcal{C}$ interaction, both agents receive a unit payoff (we set it to unity to fix the payoff scale), and in a $\mathcal{D}$-$\mathcal{C}$ interaction, the $\mathcal{D}$ receives a payoff of $b$. We set all other payoffs to zero so that a single parameter, $b$, governs the game.
In general, one may vary the two additional parameters, which correspond to the payoffs in the $\mathcal{D}$-$\mathcal{D}$ and $\mathcal{C}$-$\mathcal{D}$ interactions. We keep both these parameters equal zero for simplicity and to allow a direct comparison with Ref.\ \cite{Nowak1992}.

\textit{The rules of the Nowak-May spatial game~\cite{Nowak1992,Nowak1993}. ---} Consider a population of $L^2$ players arranged at the vertices of a regular $L\times L$ square lattice with periodic boundary conditions.  The game is played at discrete time steps, $t{=}0,1,2,\ldots, T$. The transition of the system from time $t$ to the next time step $t{+}1$ consists of the update of strategies of all players in parallel (we call this a game round). At time step $t$, an agent located at the lattice site \textbf{x} ``interacts'' with its neighbors and receives a total payoff,
\begin{equation}
P_\vx(t) = \sum_{\vy\in \mathrm{n.n.}} \vec{s}_\vx^{\,T} \widehat{H} \vec{s}_\vy \;, 
\label{eqn:payoff_NM}
\end{equation}
where $\vec{s}_\vx {\equiv} \vec{s}_\vx(t)$ is the strategy of an agent $\mathbf{x}$ at a time step $t$, and the summation runs over the eight neighbors of the agent $\mathbf{x}$ (similar to the chess king moves)%
\footnote{Refs. \cite{Nowak1992, Kolotev2018} also considered a modification of Eq.\ \eqref{eqn:payoff_NM} to include a self-interaction term, $\vec{s}_\vx^{\,T} \widehat{H} \vec{s}_\vx$. This modification does not lead to any qualitative differences. We therefore follow Ref.\ \cite{Nowak1993} and do not include self-interaction.}.
After all, payoffs are calculated, each player changes its state,
$\vec{s}_\vx(t) {\to} \vec{s}_\vx(t{+}1)$, to accept the strategy of their neighbor with the largest payoff, 
\begin{equation}
\max\{ P_\mathbf{x}(t), \max_{\mathbf{y}\in\mathrm{n.n.}}\{P_\mathbf{y}(t)\}\} \;.
\label{eqn:decision}
\end{equation}
This concludes the game round, and the game repeats with the new strategies of players. 

To fully specify the time dynamics of the game, we need to define the initial state at $t=0$.
We follow Ref.\ \cite{Nowak1992} and start the evolution with a random initial state where each player is
$\mathcal{C}$ with a probability $f_0$ and a $\mathcal{D}$ with a probability $1-f_0$. 
In the numerical simulations, we average over several realizations of initial conditions at fixed $f_0$.

Note that the initial conditions provide the only source of randomness in the game: the dynamics is fully
deterministic given the initial state of all strategies at $t{=}0$. The strategy of a
player $\vx$ at the next time step, $\vec{s}_\vx($t+1$)$, depends on the current state of
itself, $\vec{s}_\vx(t)$, its eight neighbors, and their neighbors---i.e., on states of
25 players in total. In the cellular automata language, the spatial evolutionary game of
Nowak-May (NM game) is a cellular automaton with the single-player's transition
matrix containing $2^{25}$ rules.

\textit{The steady-state of the spatial evolutionary game.---} The game starts with some initial distribution of strategies at $t{=}0$
and proceeds repeatedly. After a certain transition time, the system reaches a
steady-state, with the nature of the steady-state strongly dependent on
the value of the game parameter $b$ \cite{Nowak1992}.
Following Ref.~\cite{Nowak1992}, we characterize the instantaneous state
of the game by the density of cooperators, $f_c(t)$, and the steady-state is
characterized by the mean density of cooperators $\left<f_c\right>$:
\begin{eqnarray}
\left<f_c\right> &=&  \lim_{T\to\infty} \frac{1}{T-\tau} \sum_{t=\tau}^{T}{f_c(t)} \;, \label{eqn:mean-fc} \\
f_c(t) &=& \frac{1}{L^2} \sum_{\vx} \delta\left( \vec{s}_\vx(t) = \mathcal{C} \right)\;. \label{eqn:fc} 
\end{eqnarray}
The summation in Eq.~\eqref{eqn:fc} runs over the states of all agents at time $t$, and
$\delta\left( \vec{s}_\vx(t) {=}\mathcal{C} \right)$ is an indicator variable taking the value of 1 if $\vec{s}_\vx(t) {=}C$ and zero otherwise. In other words, Eq.~\eqref{eqn:fc} is nothing but the number of cooperators divided by the total number of agents.
In Eq.~\eqref{eqn:mean-fc}, $T$ is the total number of game rounds, and 
$\tau$ is the burn-in time such that $|\left< f_c\right> {-} f_c(t)| {\ll} \left< f_c\right>$ for $t {>} \tau$.

\textit{The mean-field modification.---} In this Letter, we include the
mean-field term as follows: instead of Eq~\eqref{eqn:payoff_NM} we consider
payoffs of the form
\begin{gather}
P_\mathbf{x}(t) = \sum_{\mathbf{y}\in \mathrm{n.n.}} \vec{s}_\mathbf{x}^{\,T} \widehat{H} \vec{s}_\mathbf{y} + %
\lambda\, \vec{s}_\mathbf{x}^{\,T} \widehat{H}_\mathrm{mf} \vec{s}_\vx \;, \label{eqn:payoff_mf_0}\\
\intertext{where}
\widehat{H}_\textrm{mf} = %
\begin{pmatrix}
f_c(t) & 0 \\
0 & b\,f_c(t) \\
\end{pmatrix} \;.
\label{eqn:payoff_mf}
\end{gather}
In other words, the interaction of a player with its neighbors is identical to the the NM game, Eq.~\eqref{eqn:payoff_NM}. Additionally, we introduce coupling to the instantaneous density of cooperators, $f_c(t)$.
This way, transition rules become non-local in space due to the dependence on the density of cooperators at the current time step $f_c(t)$. Here, we stress that $f_c(t)$ is the instantaneous density so that the game is memory-less.

In Eqs.~\eqref{eqn:payoff_mf_0}-\eqref{eqn:payoff_mf}, $\lambda$ is the
relative strength of the MF coupling. The limit $\lambda{\to} 0$ is simply the NM game,
Eq.~\eqref{eqn:payoff_NM}. 
In this work, we only consider $\lambda{=}1$, so that $b$ is the only parameter that
governs the game dynamics.

The transition from time step $t$ to $t{+}1$ is identical to the NM game: After all
payoffs are calculated, each agent changes its strategy according to
Eq.~\eqref{eqn:decision}, and the game repeats with the new strategies of players.

\begin{figure}[!htbp]
    \includegraphics[width=0.98\columnwidth, keepaspectratio=True]{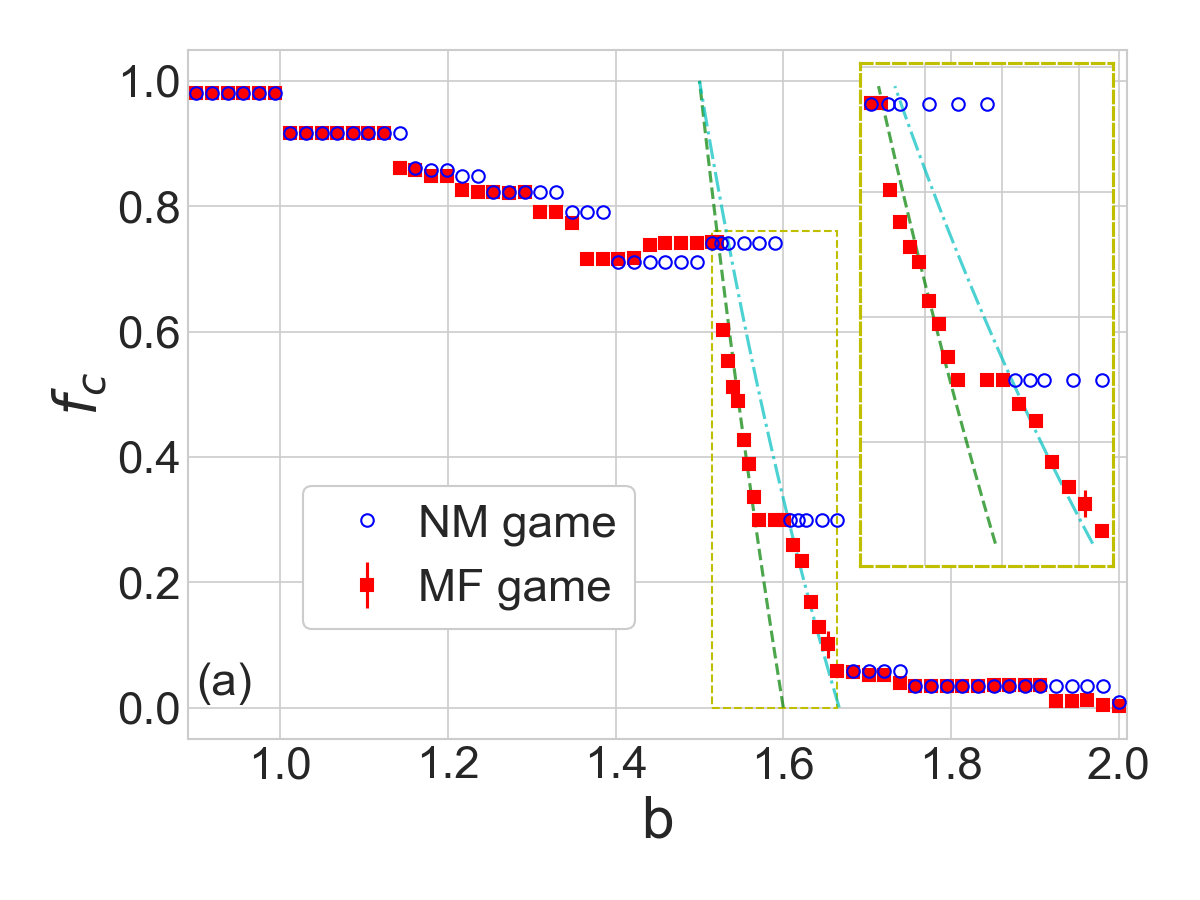}

    \includegraphics[width=0.98\columnwidth, keepaspectratio=True]{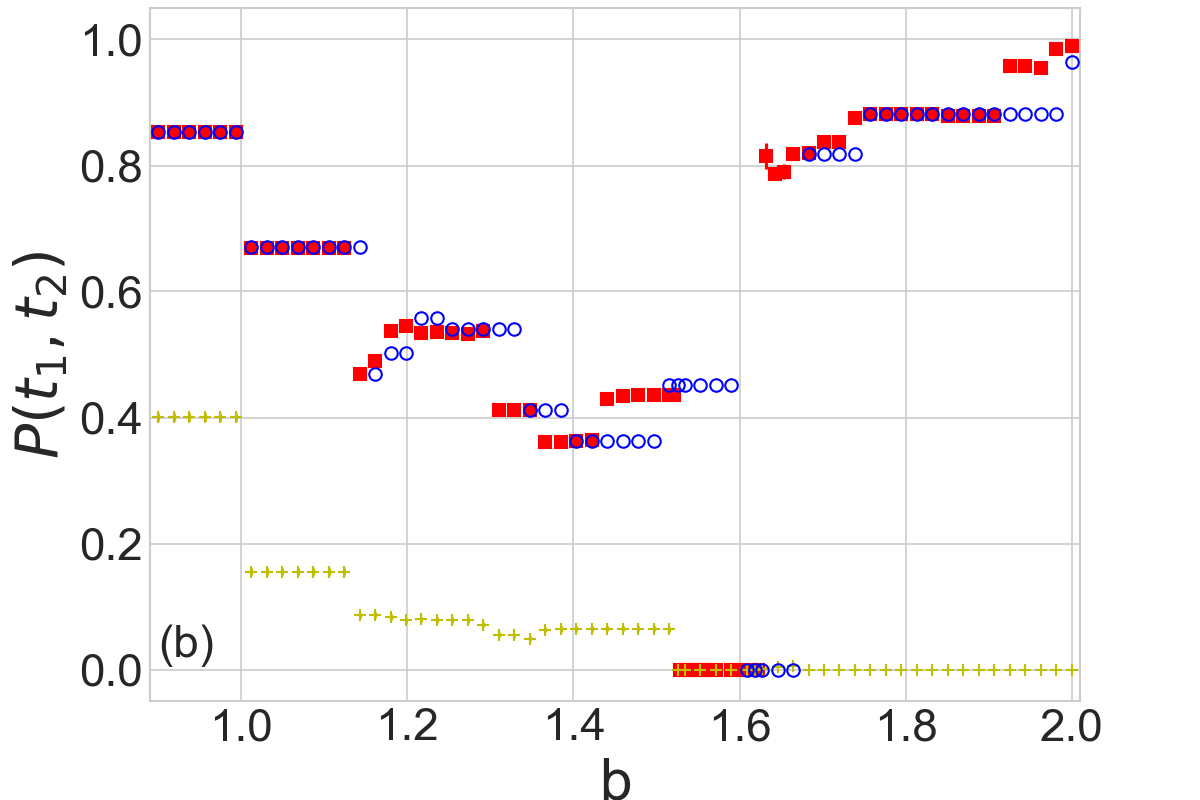}
    \caption{(a) Steady-state mean density of cooperators $\left<f_c\right>$ for the NM game (open blue circles)
    and MF game (full red squares) as a function of the payoff parameter $b$. Errorbars are shown for MF game at all points and are typically
    less than symbol size. We also show hyperbolas, Eq.~\eqref{eqn:mf_switches} for $m, n {=} (8, 5)$ (green dashed line) and 
    $m, n = (5, 3)$ (cyan dash-dotted line). In these simulations, $f_0{=}0.9$, $L=400$, $\tau{=}10^4$, $T{=}1.6{\cdot} 10^4$.
Inset shows the zoom-in of the region $1.51 {<} b {<} 1.66$. \\
    (b) Steady-state persistence, $ P(\tau, \infty)$, for the NM game (open blue circles) and the MF game (red squares).
    We also show $P(0, \infty)$ for the MF game (yellow crosses). See text for discussion.}
    \label{fig:fig1}
\end{figure}

\textit{Dynamical regimes.---} The discrete structure of the payoffs in the NM game \eqref{eqn:payoff_matrix_NM}-\eqref{eqn:payoff_NM} leads to a very specific dependence of the game dynamics on the payoff parameter $b$. Comparing the payoffs of $\mathcal{C}$ and $\mathcal{D}$ in the neighborhood of an agent, one finds for the NM game~\cite{Nowak1993, Kolotev2018} that the dynamics of the game changes at the values of $b {=} m / n$ with $1 {\leqslant} m, n {\leqslant} 8$. Here $m$ and $n$ are integers and are related to the numbers of $\mathcal{C}$ in a local neighborhood. The steady-state density of cooperators, $\langle f_c \rangle$ takes discontinuities at these special values of $b$, as shown in Fig.~\ref{fig:fig1}(a) with the open circles. 

For the mean-field game \eqref{eqn:payoff_mf_0}-\eqref{eqn:payoff_mf} (MF game),
a similar comparison of the payoffs of neighboring agents shows that an agent 
changes state depending on the ratio of $m + f_c(t)$ and $n + f_c(t)$ --- which self-consistently depends on the density of cooperators at time $t$. Since in the steady state we expect $f_c(t) {\approx} \langle f_c \rangle$, we conjecture that the steady state density is related to the payoff parameter via
\begin{equation}
b = \frac{m + \left<f_c\right>}{n + \left<f_c\right>}\;, 
\label{eqn:mf_switches}
\end{equation}
where $m$ and $n$ are integers and $1 {\leqslant} m, n {\leqslant 8}$. Note that Eq.\ \eqref{eqn:mf_switches} cannot be directly interpreted as giving the boundaries of dynamical regimes. To clarify the status of Eq.\ \eqref{eqn:mf_switches} we compare its predictions to direct numerical simulations.
   
\textit{Numerical simulations.---}
We simulate the dynamics of the game on the square lattice with linear size $L{=}400$.
We use periodic boundary conditions to minimize finite-size effects.
Additional simulations with system size up to $L{=}800$ demonstrate no significant finite-size effects.
Therefore we believe our simulations correspond to the thermodynamic limit of the model
\footnote{At smaller values of $L$ some $b$ realizations of initial conditions
contain long-lived artifacts, which disappear as the field size is increased.
}.
For all values of $b$, we simulate 100 runs with random initial distributions of $\mathcal{C}$ and $\mathcal{D}$ using the initial density of cooperator states $f_0$.
Each configuration converges to a particular steady-state, and we compute averages for both time fluctuations in the steady-state and realizations of initial conditions.
The steady-state behavior is largely insensitive to the precise value of $f_0$, and we fix $f_0{=}0.9$ in most simulation runs %
\footnote{At lower values of $f_0$ there are initial states where all $\mathcal{C}$ are isolated at $t{=}0$ and thus immediately disappear at $t{=}1$, resulting in $\left<f_c\right>{=}0$. When these initial states are discarded, the steady-state observables are independent on $f_0$.}%
, which is the value used in Refs~\cite{Nowak1993, Kolotev2018}.
The number of game rounds is typically a few thousand for the burn-in time towards the steady-state, $\tau$, cf Eq.~\eqref{eqn:mean-fc}, which is larger than the typical relaxation time for the initial state. After burn-in, we collect statistics for a few thousand further game rounds, $T{-}\tau$.

Figure~\ref{fig:fig1}(a) shows the dependence of the steady state mean density
of cooperators, $\langle f_c \rangle$, on the payoff parameter $b$ for both NM
and MF games. The values of $\langle f_c \rangle$ are essentially identical outside of the
interval $3/2 {<} b {< }5/3$. In this interval, however, the NM and MF games are
strikingly different. For the MF game,  the steady state density,
$\langle f_c \rangle$, varies continuously with $b$ --- the values of
$\langle f_c \rangle$ closely follow the hyperbolas, Eq.~\eqref{eqn:mf_switches},
with $(m,n){=}(8, 5)$ and $(5,3)$. 
This behavior is to be contrasted with the NM game, where $\langle f_c \rangle$ only changes via discontinuous jumps at $b{=}3/2$, $8/5$ and $5/3$. 

Another remarkable feature of the MF game, Fig.~\ref{fig:fig1}(a), is the flat segment  between 
the $(m, n) {=} (8,5)$ and $(5,3)$ states at $\langle f_c \rangle {\approx} 0.3$---which corresponds to the value of
$\langle f_c \rangle$ in the chaotic regime of NM game. The switch between two
regimes is reminiscent of the first-order phase transitions, driven by the
competition between two stable states. In our case, it is the switch between two
competing states---with $(m, n) {=} (8, 5)$ and $(5, 3)$---driven by the mean-field coupling.

The mean-field coupling, Eqs.\ \eqref{eqn:payoff_mf_0}-- \eqref{eqn:payoff_mf}, changes the nature of the transitions between steady state regimes: for the NM game the transitions are discontinuous; for the MF game, the density $\langle f_c \rangle$ is continuous, but its derivative w.r.t. $b$ jumps. The locations of transitions, $b_*$, are defined by the crossings of constant values $f = f_*$ and hyperbolas, Eq.\ \eqref{eqn:mf_switches}. In the vicinity of a transition, expanding  Eq.\ \eqref{eqn:mf_switches} for small $|b - b_*|$, we find
\begin{equation}
f - f_* = A_{(m, n)} |b - b_*|^\beta \;,
\label{eqn:transitions}
\end{equation}
where the amplitudes $A_{m, n} \propto 1/(m-n)$ and the pseudo-critical exponent $\beta=1$.
Table \ref{tab:amplitudes} lists the numeric values of amplitudes for transitions shown in Fig.\ \ref{fig:fig1}(a).
\begin{table}[h]
\begin{tabular}{c @{\hskip 5pt} c @{\hskip 5pt} c @{\hskip 5pt} c @{\hskip 12pt} c}
$(m, n)$ & $f_*$ & $b_*$ & $A_{(m, n)}$ \\
\hline
(8, 5)   &  0.74    &  1.522    & -10.992 & $b > b_*$\\
(8, 5)   &  0.3   &  1.566 & 9.362 &  $ b < b_*$\\
(5, 3)   &  0.3   &  1.606 & -5.444 & $b > b_*$\\
(5, 3)   &  0.23    &  1.618    & 5.232 & $b < b_*$ \\
\end{tabular}
\caption{Amplitudes for the transitions between steady states in Fig.\ \ref{fig:fig1}(a). See text for discussion. \label{tab:amplitudes} }
\end{table}

\textit{Spatial chaos.---} A salient feature of the NM game is the appearance of
the chaotic regime: for $8/5 {<} b {<} 5/3$ all agents change their strategy with
typical time scales of several game rounds while maintaining the mean density
constant, $\langle f_c \rangle {\approx} 0.3$~\cite{Nowak1993}.
(For the NM game with self-interactions, the chaotic regime occurs
for $9/5 {<} b {<} 2$ with a slightly larger density, $\langle f_c \rangle {\approx} 12\ln{2} {-} 8 {\approx} 0.318$ \cite{Nowak1992}).

The standard way of probing the chaotic behavior is via the so-called persistence~\cite{Derrida1994, Stauffer1994, Majumdar1999} $P(t_1, t_2)$, which in the present context is nothing but the fraction of players who never change their strategy between time steps $t_1$ and $t_2$. Note that we are primarily interested in the \emph{asymptotic steady state persistence}, 
$P(t_1, t_2{\to}\infty)$. To this end, we compute $P(\tau, T)$ with $\tau$ and $T$ having the same meaning as in Eq.~\eqref{eqn:mean-fc}. Averaging this value over the realizations of initial conditions, we obtain an estimate of the asymptotic value, $P(0, \infty)$.
We stress that $P(\tau, \infty)$ is qualitatively different from $P(0, \infty)$: The latter is sensitive to the overlap between the steady state and the initial distribution of strategies at $t{=}0$.

Fig.~\ref{fig:fig1}(b) compares the asymptotic persistence for the NM and MF games. Two kinds of regimes are seen for both games: those with $P(\tau, \infty) {\neq} 0$ --- where the steady-state features finite clusters (or ``cliques'') of agents who cooperatively maintain a fixed strategy,  and chaotic regimes where $P(\tau, \infty) {=} 0$ --- where everyone changes their strategy at least once. 

The dependence of $P(\tau, \infty)$ for the NM and MF games is very similar,
and the main difference is observed for $3/2 {<} b {<} 5/3$: The NM game displays
chaotic steady states for $8/5 {<} b {<} 5/3$ as expected; 
for the MF game the chaotic regime is shifted to $1.53 {\lesssim} b {\lesssim} 1.6$,
which is exactly the range where the mean density of cooperators,
$\langle f_c \rangle$, follows Eq.\ \eqref{eqn:mf_switches}.

In other words, the effect of the mean-field coupling,
Eqs.~\eqref{eqn:payoff_mf_0}-\eqref{eqn:payoff_mf}, is drastic for steady states
of the NM game, which are stable but is parametrically close
in $b$ to chaotic states: the MF coupling drives such stationary states
into spatial chaos.

We also show in Fig.~\ref{fig:fig1}(b) the behavior of $P(0, \infty)$ for the MF game. For $b {<} 1$, about 40\% of agents keep their initial strategy. Upon increasing $b$, $P(0, \infty)$ jumps to below ${\sim} 20$\% at $b{=}1$ and then further decreases before dropping off to zero at the onset of the chaotic regime. Note that $P(0, \infty)$ keeps being essentially zero for $b {>} 1.63$ where $P(\tau, \infty)$ is non-zero. This means that while the finite fraction of agents stabilize, the steady-state of an agent is not related to the initial state at $t{=}0$, which is forgotten over the first ${<}\tau$ time steps.. 

\begin{figure}[!ht]
    \includegraphics[width=0.49\columnwidth, keepaspectratio=True]{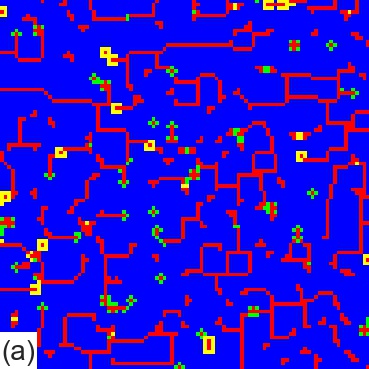}~%
    \includegraphics[width=0.49\columnwidth, keepaspectratio=True]{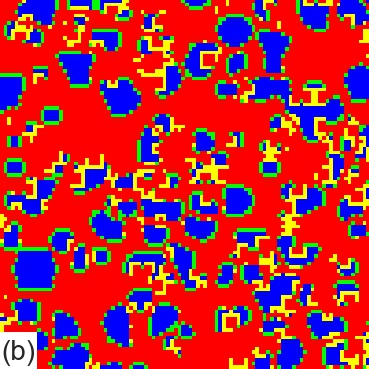}

    \includegraphics[width=0.49\columnwidth, keepaspectratio=True]{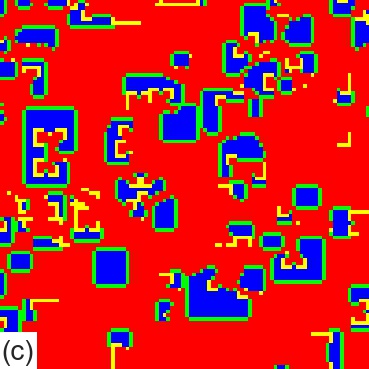}~%
    \includegraphics[width=0.49\columnwidth, keepaspectratio=True]{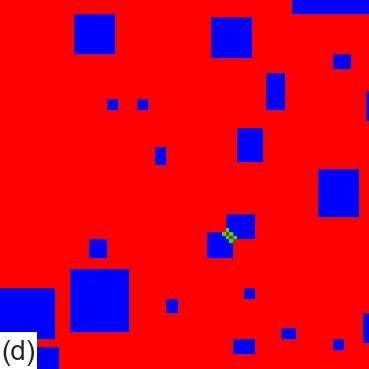}
    \caption{Typical snapshots of the steady-state configurations of players for the MF game.
    Here we show $100\times100$ part of the field with $L{=}400$, and values of $b{=}1.2$(a), $b{=}1.6$(b), $b{=}1.62$(c) and $b{=}1.63$(d).
    Blue (light grey) color corresponds to the cooperator state $\mathcal{C}$, red (dark grey) color is the defector state $\mathcal D$, green color is a $\mathcal{C}$ which was a $\mathcal{D}$ at the previous time step, and yellow is a $\mathcal{D}$ which was a $\mathcal{C}$.}
\label{fig:fig2}
\end{figure}

\textit{Game field configurations.---} Fig.~\ref{fig:fig2} shows
representative snapshots of steady states of the game field 
of the MF game depending on the payoff parameter $b$. For $b {<}1.53$ typical configurations
feature web-like structures of $\mathcal{D}$ players. These webs are random,
i.e., precise positions of the branches vary depending on an initial state
of the field(see Fig.~\ref{fig:fig2}(a)). In the steady state, these webs remain
mostly static, with possible ``blinking'' players at the interfaces. The widths
of the branches of defectors increases with increasing the payoff parameter $b$,
and at around $b{\approx} 1.53$---which is the onset of the chaotic regime--- the
$\mathcal{D}$ webs melt into collections
of smaller clusters of $\mathcal{C}$ and $\mathcal{D}$, which are no longer
static and grow, shrink and collide instead.

At $b{>}1.53$ field is dominated with clusters of $\mathcal{C}$ and $\mathcal{D}$ of
varying shapes and sizes (Fig.~\ref{fig:fig2}(b)).
Upon further increase of the payoff parameter, for $1.6 {<} b {<}1.62$ the field is dominated by square-shaped
clusters of $\mathcal{C}$(see Fig.~\ref{fig:fig2}(c)). Isolated clusters grow in all directions, and
disintegrate upon colliding with neighboring clusters. For $b {>} 1.63$ the
fields feature static collections of small square-shaped clusters of
$\mathcal{C}$ embedded into the $\mathcal{D}$ background(see Fig.~\ref{fig:fig2}(d)).

The rectangular shapes of spatial patterns that emerge in Fig.\ \ref{fig:fig2} can be readily understood from the synchronous evolution rules and the payoff structure of the NM game. Consider an interface between regions of $\mathcal{C}$ and $\mathcal{D}$ agents. The interface can be either straight (i.e., oriented along the square lattice directions), or ``angled'' (e.g., oriented along the diagonal of the lattice). For the NM game ($\lambda=0$ in Eq. \eqref{eqn:payoff_mf}), a square cluster of $\mathcal{C}$ is stable for $5/3 < b < 8/3$; however the ``angled'' interface is only stable for $7/3 < b < 8/3$.  For $b < 7/3$, the angled interface evolves into a straight one in several time steps---and the straight interface is stable. Introducing the mean-field term, Eq.\ \eqref{eqn:payoff_mf} changes  switch-over ratios, $m/n$, into density-dependent hyperbolas, Eq.\ \eqref{eqn:mf_switches}.

\begin{figure}[!ht]
\includegraphics[width=0.49\columnwidth, keepaspectratio=True]{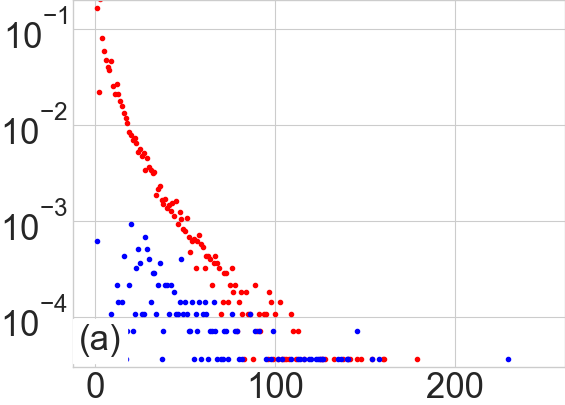}~%
\includegraphics[width=0.49\columnwidth, keepaspectratio=True]{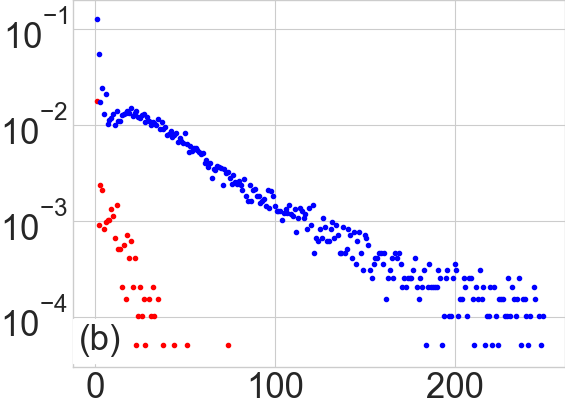}
    
\includegraphics[width=0.49\columnwidth, keepaspectratio=True]{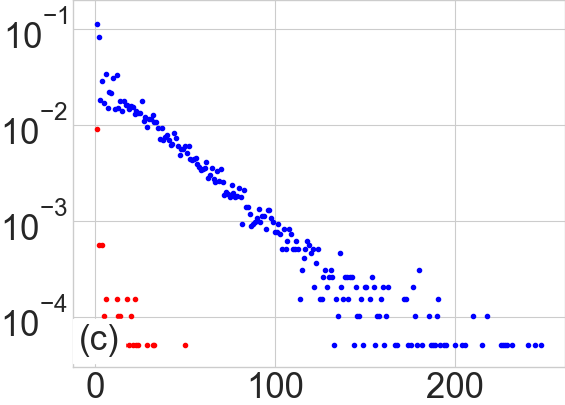}~%
\includegraphics[width=0.49\columnwidth, keepaspectratio=True]{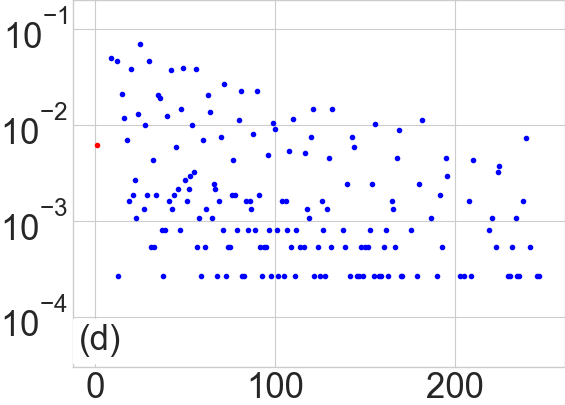}
\caption{Distribution of the cluster sizes (i.e. the number of agents in a cluster) of $\mathcal{D}$ (red) and $\mathcal{C}$ (blue) 
in the steady states for $L{=}200$, and values of $b{=}1.2$ (a), $b{=}1.55$ (b), $b{=}1.59$ (c) and $b{=}1.63$ (d). 
We only show clusters of up to 250 agents.}
\label{fig:fig3}
\end{figure}

It is instructive to examine the distribution of the sizes of clusters
of $\mathcal{C}$ and $\mathcal{D}$ in different steady-state regimes.
Figure~\ref{fig:fig3} shows the exponential decay in the distribution
of the cluster sizes in chaotic regimes, $b{=}1.55$ and $1.59$, while it is
slower than exponential for
$b{=}1.2$, the regime with the long memory of initial states
(see Fig.~\ref{fig:fig1}(b)).

Asymptotically, the fractal dimension of the clusters as well as of 
the interfaces between clusters of cooperators $\mathcal C$ and defectors
$\mathcal D$ tends to the dimension of the plane, $d{=}2$, which also
coincides with the scaling dimension of masses of clusters~\cite{Kolotev2018}.
These interfaces form a special type of random fractals, quite different from
those at the second-order or first-order phase transitions~\cite{Bunde1996}.

\textit{Discussion.---} We propose a way of introducing self-consistent mean-field-like
couplings into evolutionary games with spatially structured populations. We perform
direct numerical simulations of a simplest yet non-trivial model---where decision
rules include non-local information due to the mean-field coupling---and find a variety
of steady states: chaotic states and self-organized stationary states where a
finite fraction of agents form stable clusters (or ``cliques''). These states are
not critical because cluster sizes are distributed exponentially. 

We find that the mean-field coupling induces qualitative changes for stationary
steady states, which are parametrically close to chaotic states.

In this work, we only consider a restricted version of the Prisoner's Dilemma, 
Eq.\ \eqref{eqn:payoff_matrix_NM}. Our approach however, can be readily generalized
to a wide class of games, including PD games with non-zero $\mathcal{C}$-$\mathcal{D}$ and $\mathcal{D}$-$\mathcal{D}$ payoffs and versions of the Public Goods game. 

The introduction of the mean-field type couplings opens several intriguing avenues
for possible future work. It would be interesting to clarify the effect of the
strength of the MF coupling, $\lambda{\neq} 1$. Non-uniform, site-dependent
couplings (e.g., $\lambda_\vx$ drawn from some random distribution) would
reflect natural variations between individuals. The role of the local
connectivity of the population needs to be clarified: this includes both
higher-dimensional regular lattices and disordered and hierarchical random
lattices and graphs. One may introduce an external field. It would be
interesting to see the interplay between the mean-field couplings and noisy
decision rules (e.g., where the decision rules contain
pseudo-temperature~\cite{Hauert2005, Szolnoki2005, Javarone2016}).

\begin{acknowledgments}

L.S. is supported within the framework of State Assignment of Russian Ministry
of Science and Higher Education. E.B. and A.D. acknowledge support within the Project Teams framework of MIEM HSE. We thank J.J.~Arenzon and A.~Szolnoki for
valuable suggestions. Simulations were carried out in part
through computational resources of HPC facilities at HSE University.

\end{acknowledgments}


\end{document}